# Retrieving the quantitative chemical information at nanoscale from SEM EDX measurements by Machine Learning

B.R. Jany[*], A. Janas, F. Krok

Marian Smoluchowski Institute of Physics Jagiellonian University, Lojasiewicza 11, 30-348 Krakow, Poland

**Abstract**

The quantitative composition of metal alloy nanowires on InSb(001) semiconductor surface and gold nanostructures on germanium surface is determined by blind source separation (BSS) machine learning (ML) method using non negative matrix factorization (NMF) from energy dispersive X-ray spectroscopy (EDX) spectrum image maps measured in a scanning electron microscope (SEM). The BSS method blindly decomposes the collected EDX spectrum image into three source components, which correspond directly to the X-ray signals coming from the supported metal nanostructures, bulk semiconductor signal and carbon background. The recovered quantitative composition is validated by detailed Monte Carlo simulations and is confirmed by separate cross-sectional TEM EDX measurements of the nanostructures. This shows that SEM EDX measurements together with machine learning blind source separation processing could be successfully used for the nanostructures quantitative chemical composition determination.

**Keywords:** SEM, EDX, Machine Learning, BSS

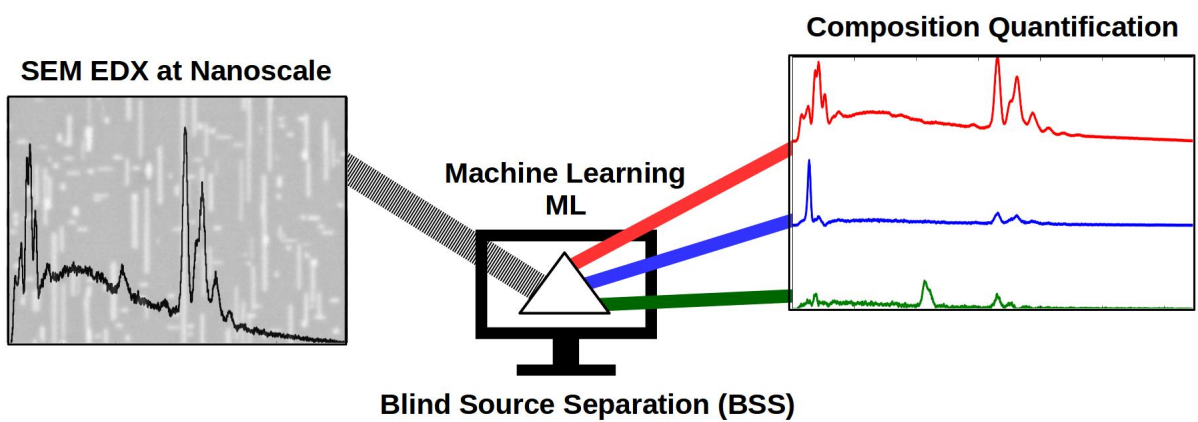

* corresponding author e-mail: benedykt.jany@uj.edu.pl

The Scanning Electron Microscope (SEM) with a Field Emitter Gun (FEG) electron source became a popular tool for the nanoscience[1]. It can deliver the information at the nanoscale on the sample topography by collecting the secondary electrons (SE) and relative sample composition by the backscattered electrons (BSE), which emission is related to the mean atomic number. It is also very common that a SEM is equipped with energy dispersive X-ray spectroscopy (EDX) system. Nowadays such a EDX system usually consist of high efficient Silicon Drift Detector (SDD) capable of recording high count rates. The spatial resolution in the SEM EDX mapping is related to the interaction volume of primary electron beam and consequently X-ray generation volume. Careful optimization of the X-ray depth distribution and spatial radial distribution by adjusting the electron beam energy and size (the beam current) leads to the acquisition of high spatial resolution X-ray maps at nanoscale[2; 3]. However the quantification of the recorded SEM EDX from nanostructures is challenging due to the mixing of the signals from different depths of the sample, resulted from X-ray generation depth. This is very similar as for the TEM EDX for the heterogeneous volumes, where there is a spatial overlap of the different phases in the beam path[4] . For the separation of the components from the mixture Machine Learning (ML), methods such as blind source separation (BSS) using independent component analysis (ICA)[5]  and non negative matrix factorization[6]  (NMF), are successfully applied. As shown it work for the TEM EELS measurements[7; 8]  and recently for TEM EDX measurements of multicomponent signal unmixing of nanoheterostructures[4; 9]. The idea of BSS method is to statistically decompose the mixed signal into separate sources, without any external information. These methods are widely used also in the different fields of science[10-12].  Here we apply the BSS decomposition using NMF to SEM EDX spectrum image maps of metal alloy nanowires grown on AIIIBV semiconductor surface. The number of decomposition components is provided by principal component analysis (PCA). The quantitative composition of nanowires is recovered, the results of the quantification are additionally verified by detailed Monte Carlo simulations. The nanowires composition is confirmed by separate cross-sectional TEM EDX measurements.

The $AuIn_2$ metal alloy nanowires on InSb(001) (AIIIBV semiconductor) surface were prepared by molecular beam epitaxy (MBE) deposition of 2 mono-layers (ML) of gold on atomically clean reconstructed InSb(001) surface at temperature of 330C in ultra high vacuum conditions (UHV). Such a perpetration conditions results in the formation of $AuIn_2$ metal alloy nanowires on the surface in the process of thermally induced self-assembly[13] . The AIIIBV semiconductors since their unique properties are seriously considered for future electronic devices especially that the technology to integrate the AIIIBV at the nanoscale with silicone [14; 15] was developed. The gold-rich nanostructures on AIIIBV semiconductors are widely used as a catalyst to grow standing arrays of vertically aligned AIII-BV nanowires[16; 17] for many applications as for example efficient water reduction[18] or nano light emitting diodes (LED) with high brightness[19] . They also have a potential usage as nanoelectrodes and ohmic contacts[20] . Similarly the Au hcp nanostructures, gold of rare and unique hexagonal structure, were prepared on Ge(001) surface as we recently shown[21; 22] . These have a potential usage as a bridge connecting the existing cubic semiconductors like germanium with hexagonal ones like boron nitride. The Au/Ge(001) surface itself is also very interesting for electronic applications due to the existence of 1D and 2D conduction channels in form of atomic chains[23] and subsurface layer[24] . After samples preparation in UHV the surface of the samples was covered by thermally evaporated carbon capping layer to prevent surface damage and oxidation.

The SEM EDX data were acquired using Double Beam SEM/FIB Quanta 3D FEG microscope by FEI equipped with EDAX Ametek Apollo XPP SDD EDX detector with an active area of 10mm$^2$. The data were collected in the form of spectrum image (SI) where at each pixel the full EDX spectrum was collected during sample surface scanning by SEM electron probe. Simultaneously the BSE image was collected, by 4-quad semiconductor BSE detector mounted at the pole piece of electron column, where the intensity of collected BSE electrons is proportional to the average atomic number. The EDX data were measured in the form of 3D stack, where for each x, y sample grid point a full EDX spectrum was collected at z axis. The 6.5keV(for AuIn2/InSb) and 6keV(for Au/Ge) electron energy was used, with a beam current of 16nA and 500us dwell time per pixel. The EDAX Genesis software from the system manufacturer was used for data acquisition and for ZAF standardless method of spectra atomic fractions composition quantification. The free software HyperSpy[25] was used for BSS data processing, the NMF and PCA was performed using algorithms as implemented in HyperSpy from Scikit learn[26] with Poisson noise normalization[27]. The free DTSA2 software from NIST [28] was used for the precise Monte Carlo SEM EDX spectra simulations.

Additional TEM EDX measurements of the FIB prepared samples cross-sections were performed using 200keV electron energy by FEI Tecnai Osiris TEM microscope equipped with Super-X EDX detector setup and high brightness Schottky X-FEG electron source. The composition quantification of the collected X-ray spectra was performed using the Cliff-Lorimer method by dedicated software ESPRIT from Bruker.

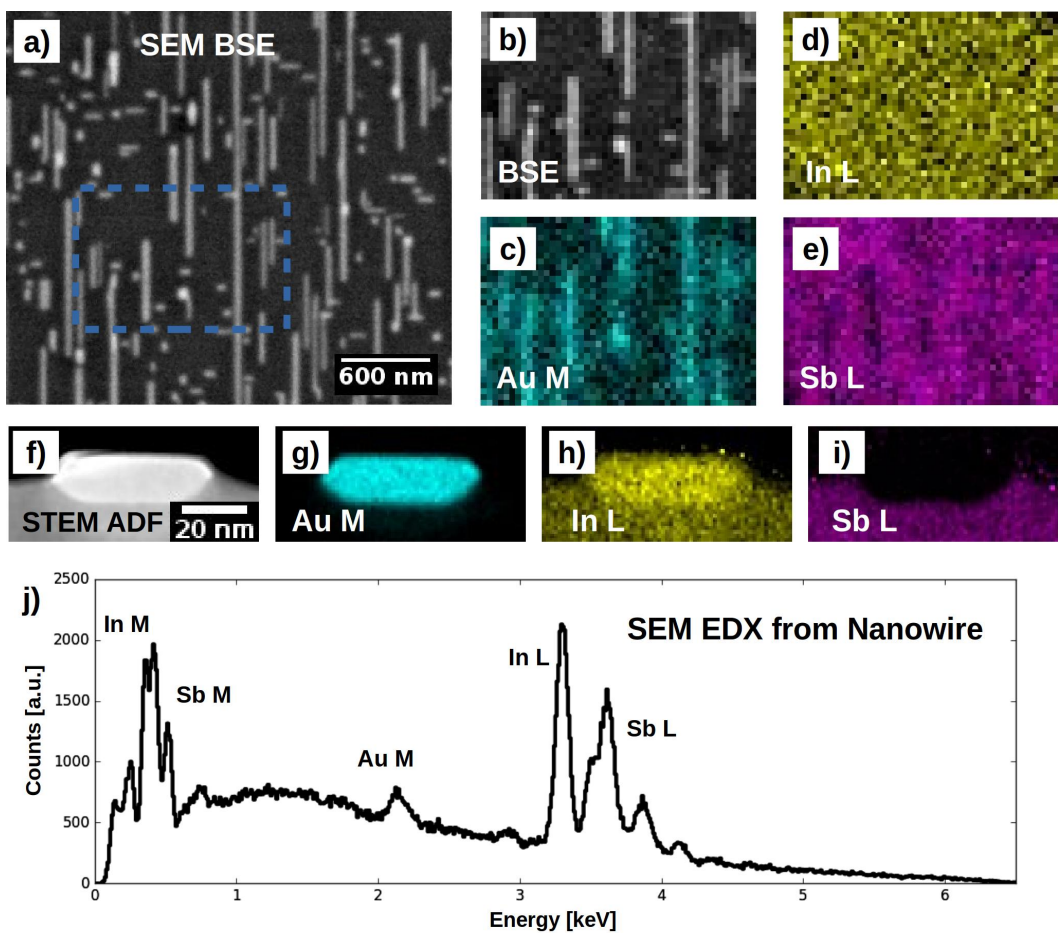

*Figure 1: AuIn2 nanowires on InSb(001) surface. a) SEM BSE image of nanowires, b) EDX analysis area. SEM EDX elemental maps of c) gold, d) indium, e) antimony. TEM EDX nanowires cross-section ADF STEM f) and elemental maps of g) gold, h) indium, i) antimony. SEM EDX spectrum from nanowire j), due to the X-ray interaction volume there is also signal from InSb below the wire.*

*AuIn2 nanowires on InSb(001).* Figure 1a-b shows SEM BSE signal from AuIn2 nanowires formed on InSb(001) surface collected during EDX spectrum image collection. The nanowires are of an average width of ~70nm and an average length of ~500nm as estimated from SEM measurements (see supplementary Fig. S1). The SEM EDX elemental maps of gold, indium and antimony Fig. 1c)-e), are extracted from X-ray intensity measurements by background subtraction. These elemental maps shows a spatial distribution of different elements, in particular the gold EDX map Fig. 1c) nicely shows the location of the nanowires. The quantitative information on the nanowires chemical composition, by examining the SEM EDX spectrum from nanowires area (Fig. 1j), cannot be obtained due to the presence of the InSb signal in the spectrum which comes from below the wires. This is due to the X-ray generation volume, as described by X-ray depth and lateral distribution (for the CASINO[29] simulations see supplementary Fig.S2-S3). The SEM EDX signal for the examined system comes from a depth of approximately 200nm. Thus the X-rays signal from AuIn2 nanowires is mixed together with InSb signal. For comparison TEM EDX cross-section ADF STEM Fig. 1f) and corresponding EDX elemental maps Fig. 1g)-i), are shown for which the X-ray signals are nicely spatially separated. The quantification of TEM EDX spectra confirms the AuIn2 nanowires stoichiomet (see supplementary Fig. S4).

The collected EDX spectrum image as in Figure 1c)-e) was subsequently processed by BSS using HyperSpy. First we performed the dimensionality reduction by PCA to determine the number of components. The results of the PCA are presented as a scree plot, the proportion of the variance for the given principal component Fig. 2a). The scree plot show that first three principal components (PC1, PC2, PC3) have the significantly higher variance then the remaining components. Next we used NMF to unmix the EDX data assuming three components present, as derived from PCA. The non negative matrix factorization (NMF) assumes that the non negative signal is a mixture of the non negative sources. When NMF is applied to such data type, the discovered components often correspond remarkably well to those sources, as noted by the review on the NMF[30]. The NMF decomposition results in three component maps (NMF1, NMF2, NMF3) Fig. 2b)-d), which show the spatial distribution of the phases, and corresponding component spectra containing X-ray lines of the elements Fig. 2f). We see that the NMF1 contains only the indium and antimony X-ray peaks (InSb phase), NMF2 contains mostly the carbon X-ray peak, originating from carbon capping, while the NMF3 contains only the gold and indium X-ray peaks (AuIn2 nanowires). In NMF2 there is also some indium signal visible, which most probably originates from the secondary fluorescence of indium by antimony. To determine the quantitative composition of the BSS decomposed phases we used ZAF method as implemented in the EDAX Genesis software. The NMF1, and NMF3 component spectra were imported into EDAX Genesis and atomic fractions composition quantification was performed by ZAF method. The results of the obtained chemical composition quantification are presented in Table 1. It is seen that the results of the NMF1(InSb) and NMF3(AuIn2 nanowires) quantification corresponds, within estimated uncertainties, to the true composition of InSb and AuIn2 phases. In order to validate the results of the NMF1 and NMF3 component spectra ZAF quantification we performed detailed Monte-Carlo simulations of X-rays collected by SDD detector by DTSA2 software. We compared the NMF1 and NMF3 component spectra with Monte-Carlo simulation of pure InSb and pure AuIn2 phase, respectively, as depicted in Fig. 3 (the spectra were normalized to the highest peak for comparison).

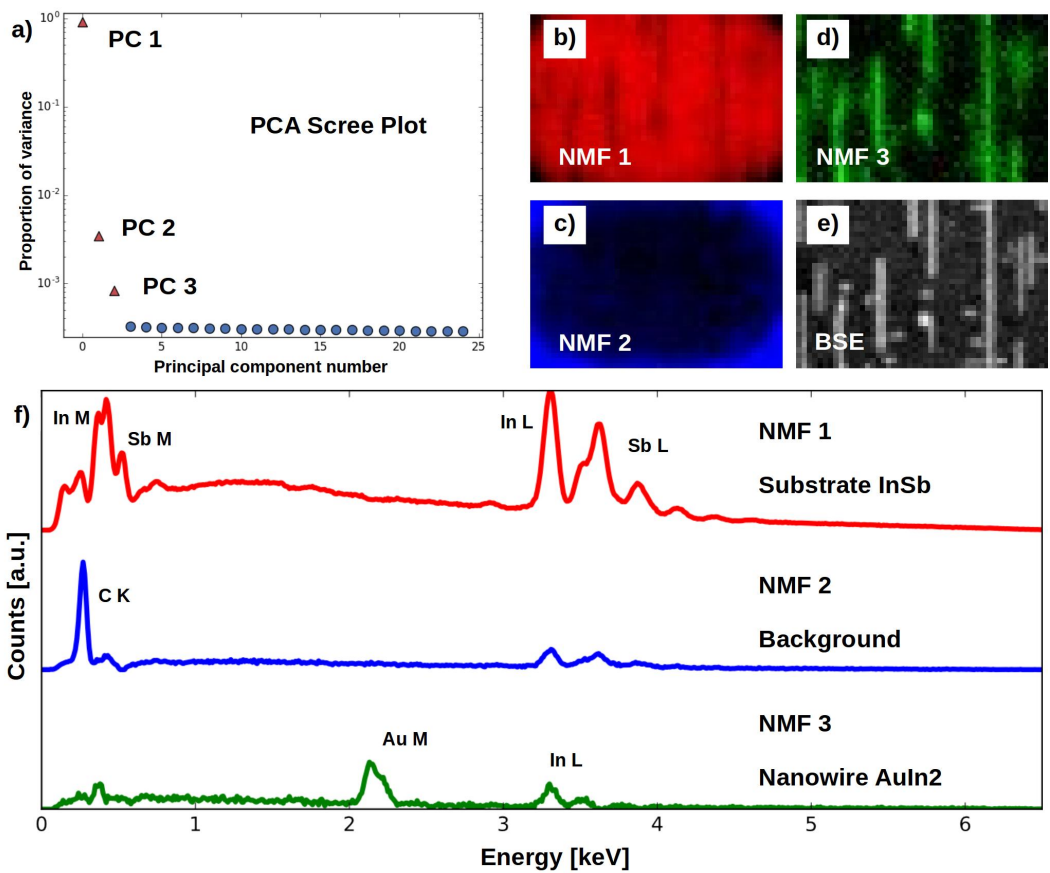

Figure 2: Results of the BSS by NMF and PCA of EDX spectrum image of AuIn2 nanowires on InSb. a) scree plot of first 25 principal components, three components exhibit significantly higher variance, result of PCA. Non negative matrix factorization component maps b)-c) and corresponding component spectra containing X-ray lines of the elements f). BSE image of the area of analysis e). The BSS by NMF separates very good the signal from AuIn2 nanowires (NMF 3) from InSb substrate (NMF 1) and from background (NMF 2).

|  | NMF 1 (InSb substrate) | NMF 3 (AuIn2 nanowires) |
|---|---|---|
| **EDX ZAF Quantification [atomic %]** | In: 45.0(4.7)   Sb: 55.0(5.7) | Au: 38.0(5.9)   In: 62.0(5.9) |
| **True Composition [atomic %]** | In: 50.00   Sb: 50.00 | Au: 33.33   In: 66.67 |

Table 1: Results of the composition quantification of BSS decomposed component spectra NMF1(InSb) and NMF3(AuIn2 nanowires)

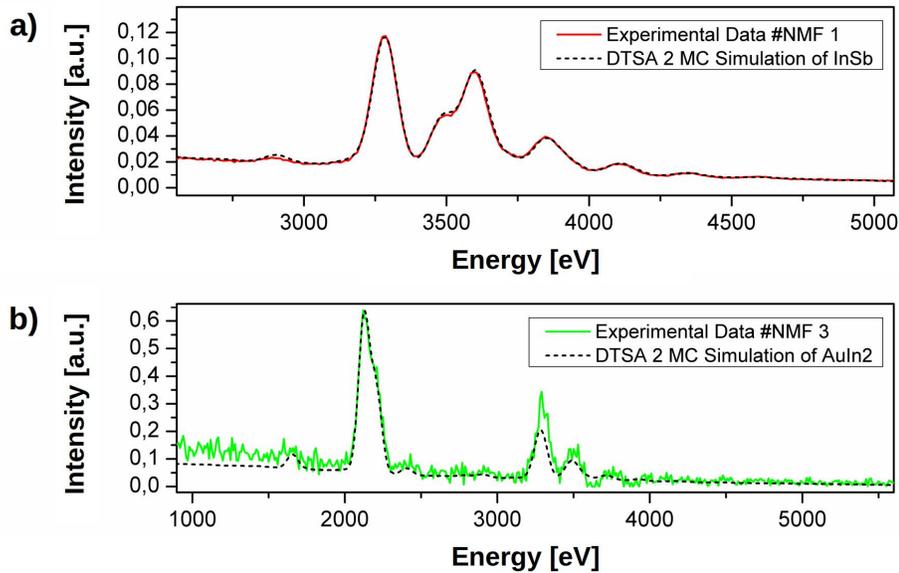

*Figure 3: Comparison of the experimentally determined BSS decomposition component spectra NMF1 a) and NMF2 b) with Monte-Carlo simulation of pure InSb and pure AuIn2 phase respectively as calculated by DTSA2 software. The simulations describe correctly all of the features of the experimental data EDX spectra.*

We see that the NMF1 spectrum matches almost perfectly the simulated InSb EDX spectrum. Also the NMF3 spectrum matches well the simulated AuIn2 EDX spectrum. In this case the overall data statistics is much smaller as seen by the background fluctuations in the spectrum around energy of 1000eV, which values are approximately ~0.1 in this scale. Nevertheless the simulations describe correctly all of the features of the experimental data EDX spectra, together with proper peak intensity scaling, within the limitation of the data statistical fluctuations. This proofs the validity of the of the performed composition quantification. In this case the BSS separated correctly the EDX signal of the AuIn2 nanowires, which was successfully quantified by ZAF method, as validated. It is also important to note that the analysis is performed without any external input about the sample composition or background, the method blindly decomposed the signal into components which reflects the phases present in the sample.

*Au hcp nanostructures on Ge(001).* We now go on with the analysis of the Au hcp nanostructures formed on Ge(001) surface. The gold nanostructures are of unique hcp phase, as we recently showed by the atomically resolved HAADF STEM measurements[22]. To distinguish whetever the nanostructures are made of pure gold or gold/germanium alloy we imployed the STEM measurements [22]. Fig. 4a) shows the SEM BSE image of grown Au hcp nanostructures. The nanostructures have an average size of ~50nm. The SEM EDX elemental maps of gold and germanium Fig. 4c)-d) show that the nanostructures are Au rich and there is also a germanium signal reduction at the position of the nanostructure. But due to the X-ray generation volume one cannot exclude the germanium content in the nanostructure (see supplementary Fig. S5 and Fig. S6). This once more again prohibits us from the composition quantification i.e. whether we are dealing here with pure gold or gold/germanium alloy nanostructure.

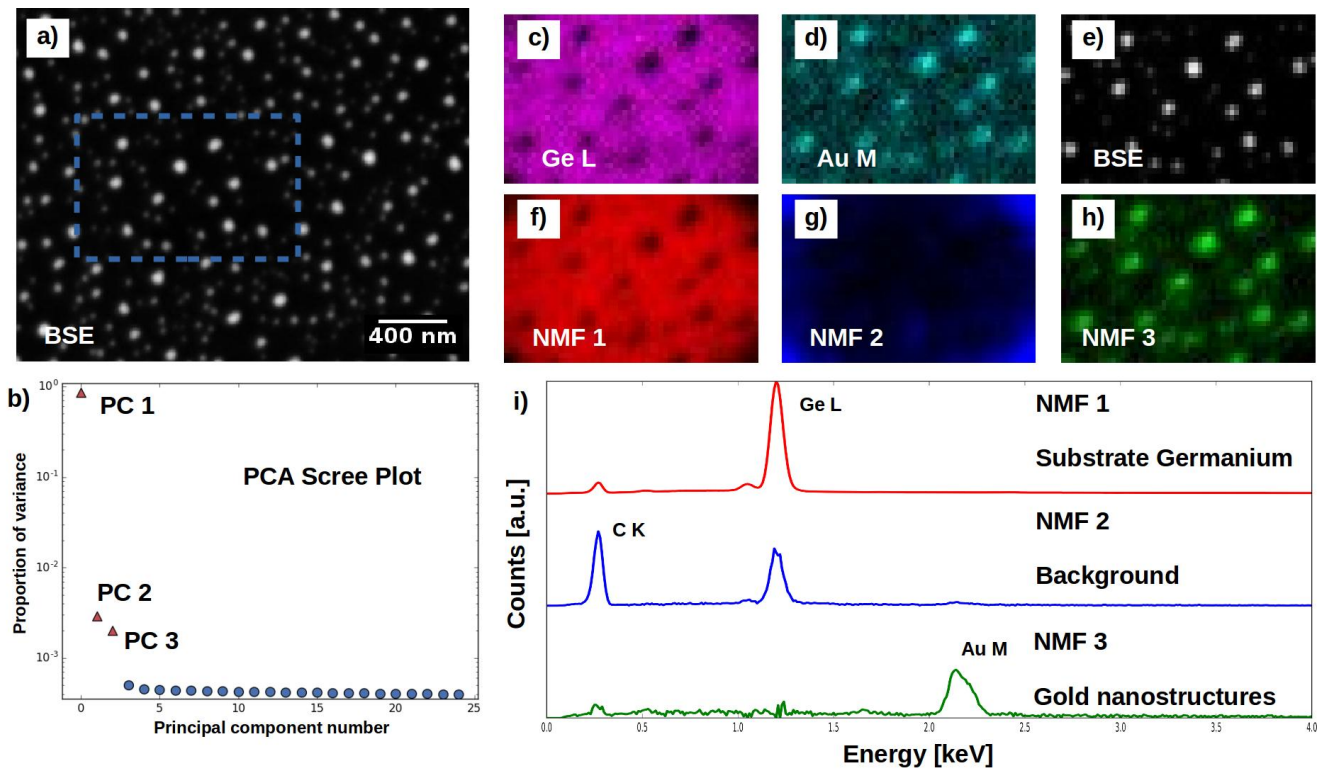

*Figure 4: Au hcp nanostructures on Ge(001) surface. a) SEM BSE image of the Au hcp nanostructures. b) PCA scree plot of first 25 principal components, three components exhibit significantly higher variance  Conventional EDX maps of germanium c) and gold d). Non negative matrix factorization component maps f)-h) and corresponding component spectra containing X-ray lines of the elements i). BSE image of the area of analysis e). The BSS by NMF once more separates very good the signal from Au hcp nanostructures (NMF 3) from Ge substrate (NMF 1) and from background (NMF 2).*

Now we perform the BSS on the same EDX spectral image using the same approach as for AuIn2 nanowires. First we performed the PCA, the scree plot Fig. 4a) shows that we have three phases in the data. As before the NFM decomposition shows three component maps Fig. 1f)-h) and corresponding component spectra Fig. 1i). The NMF1 component shows the germanium X-ray intensity and is related to the germanium phase from the bulk. The NMF2 component shows the X-ray intensity from carbon, as used for sample surface protection, and additionally some germanium signal, originating from secondary fluorescence of germanium fluorescence by gold. The NMF3 component consist of only gold X-ray intensities, which originate from gold nanostructures. By employing the BSS we can now undoubtly say that the formed nanostructures are formed only from gold, the formation gold/germanium alloy is excluded. This results are in agreement with our STEM measurements [22].

Based on the two studied examples of AuIn2 nanowires on InSb surface and Au hcp nanostrcuures on germanium surface, we have shown that by using the blind source separation techniques on SEM EDX spectral images we can successfully extract the nanostructures pure X-ray signal from other X-ray signal present in the data like bulk matrix, carbon background or secondary fluorescence. The extracted X-ray signal originating from nanostructures is used to determine the true chemical composition of the formed structures by the ZAF method, the quantification is verified by detailed Monte-Carlo simulations. The samples composition is additionally verified by cross-sectional TEM measurements. The SEM EDX spectral images measurements with the application of the blind source separation techniques could be now successfully applied for the chemical composition quantification at the nanoscale.

# Acknowledgments

Supports by the Polish National Science Center (DEC-2015/19/B/ST5/01841) is acknowledged. Part of the research was carried out with equipment purchased with financial support from the European Regional Development Fund in the framework of the Polish Innovation Economy Operational Program (Contract No. POIG.02.01.00-12-023/08). B.R.J. acknowledges the support of Polish Ministry of Science and Higher Education under the grant 7150/E-338/M/2016.

# Competing financial interests

The authors declare no competing financial interests.

# Supplementary Materials

## Retrieving the quantitative chemical information at nanoscale from SEM EDX measurements by Machine Learning

B.R. Jany[*], A. Janas, F. Krok

Marian Smoluchowski Institute of Physics Jagiellonian University in Krakow, Lojasiewicza 11, 30-348 Krakow, Poland

---

* corresponding author e-mail: benedykt.jany@uj.edu.pl

# AuIn2 nanowires size distributions from SEM measurements

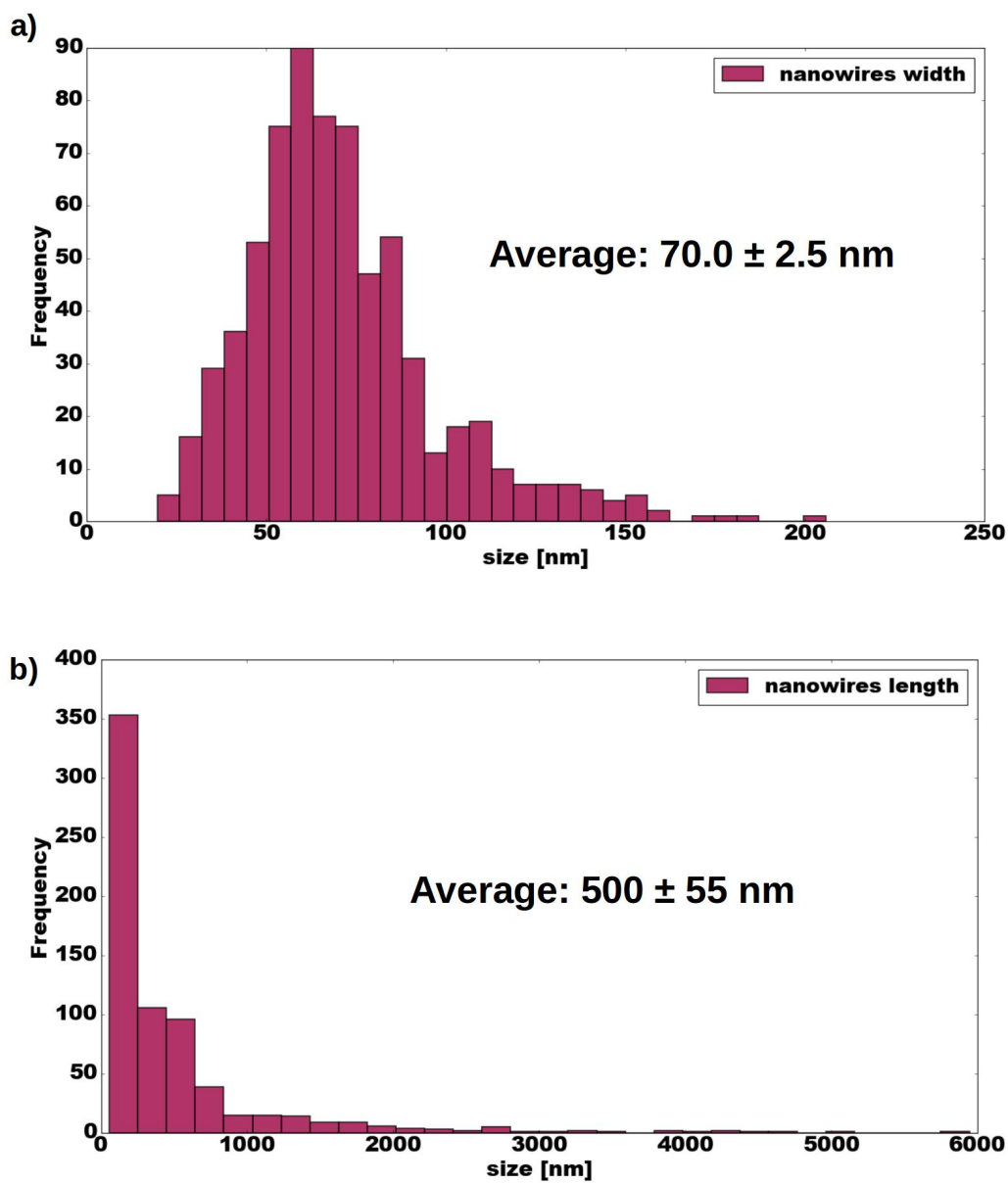

*Figure S1: AuIn2 nanowires width distribution a) and length distribution b) from SEM measurements.*

# Monte-Carlo simulation of electrons interaction by CASINO for AuIn2 nanowires on InSb

## Pure InSb

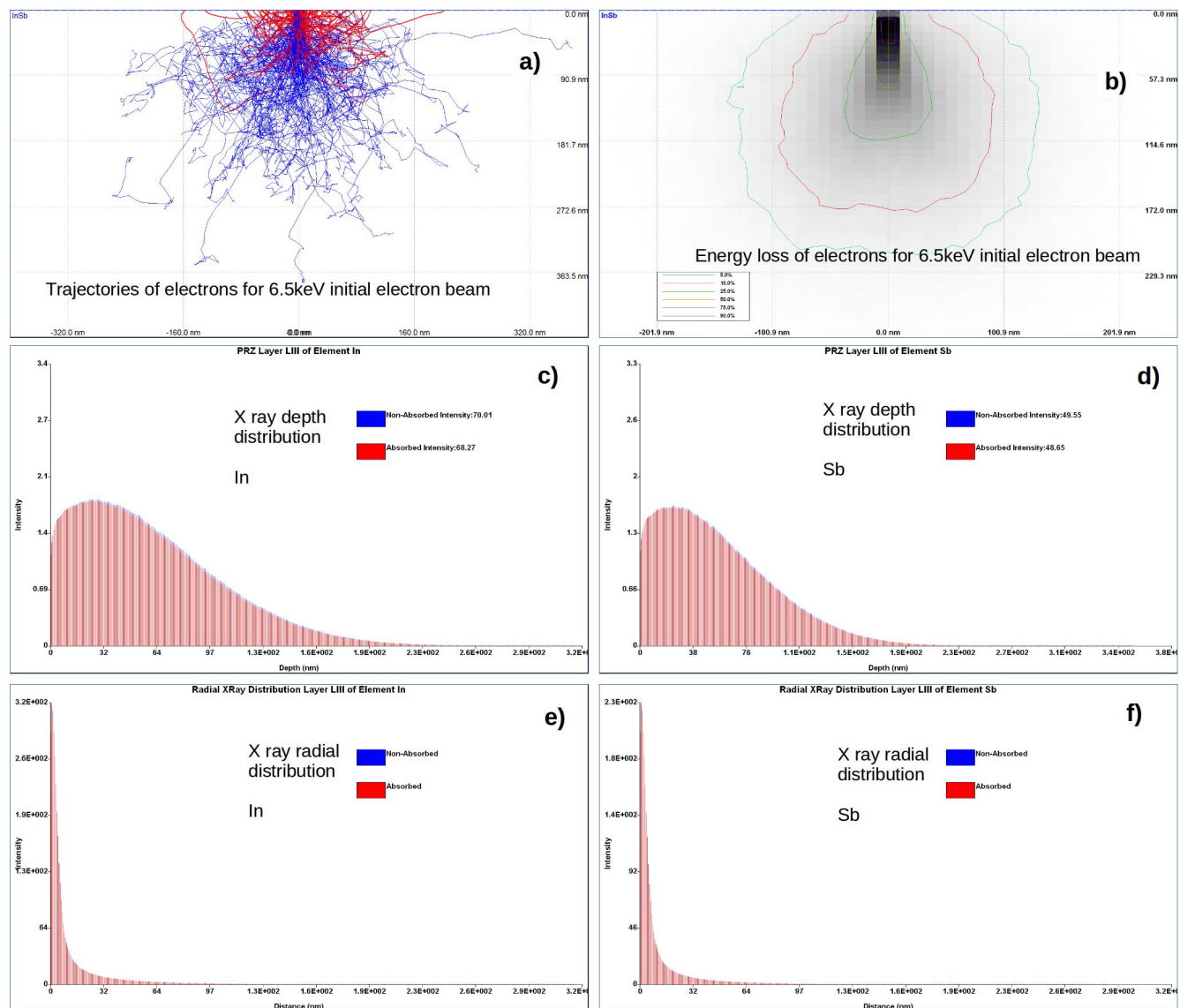

*Figure S2: Monte-Carlo simulation of 6.5keV electrons interaction with InSb by CASINO. a) Trajectories of generated electrons, b) energy loss of electrons. Generated X rays depth distribution for indium c) and antimony d). Generated X rays radial distribution for indium e) and antimony f).*

# AuIn2 nanowire on InSb

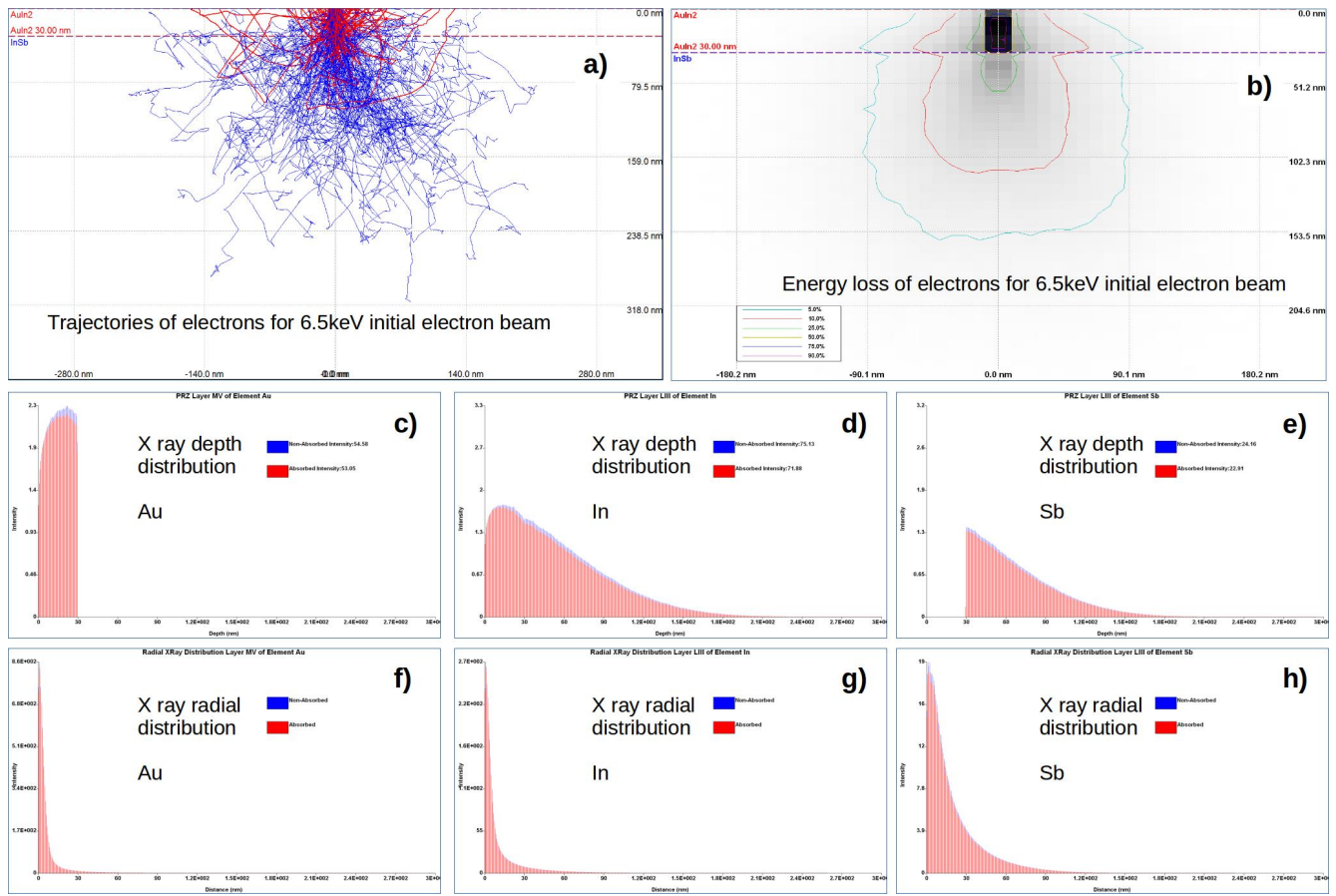

*Figure S3: Monte-Carlo simulation of 6.5keV electrons interaction with AuIn2(30nm) nanowire on InSb by CASINO. a) Trajectories of generated electrons, b) energy loss of electrons. Generated X rays depth distribution for gold c), indium d) and antimony e). Generated X rays radial distribution for gold f), indium g) and antimony h).*

# AuIn2 nanowire on InSb(001) TEM cross-section measurements

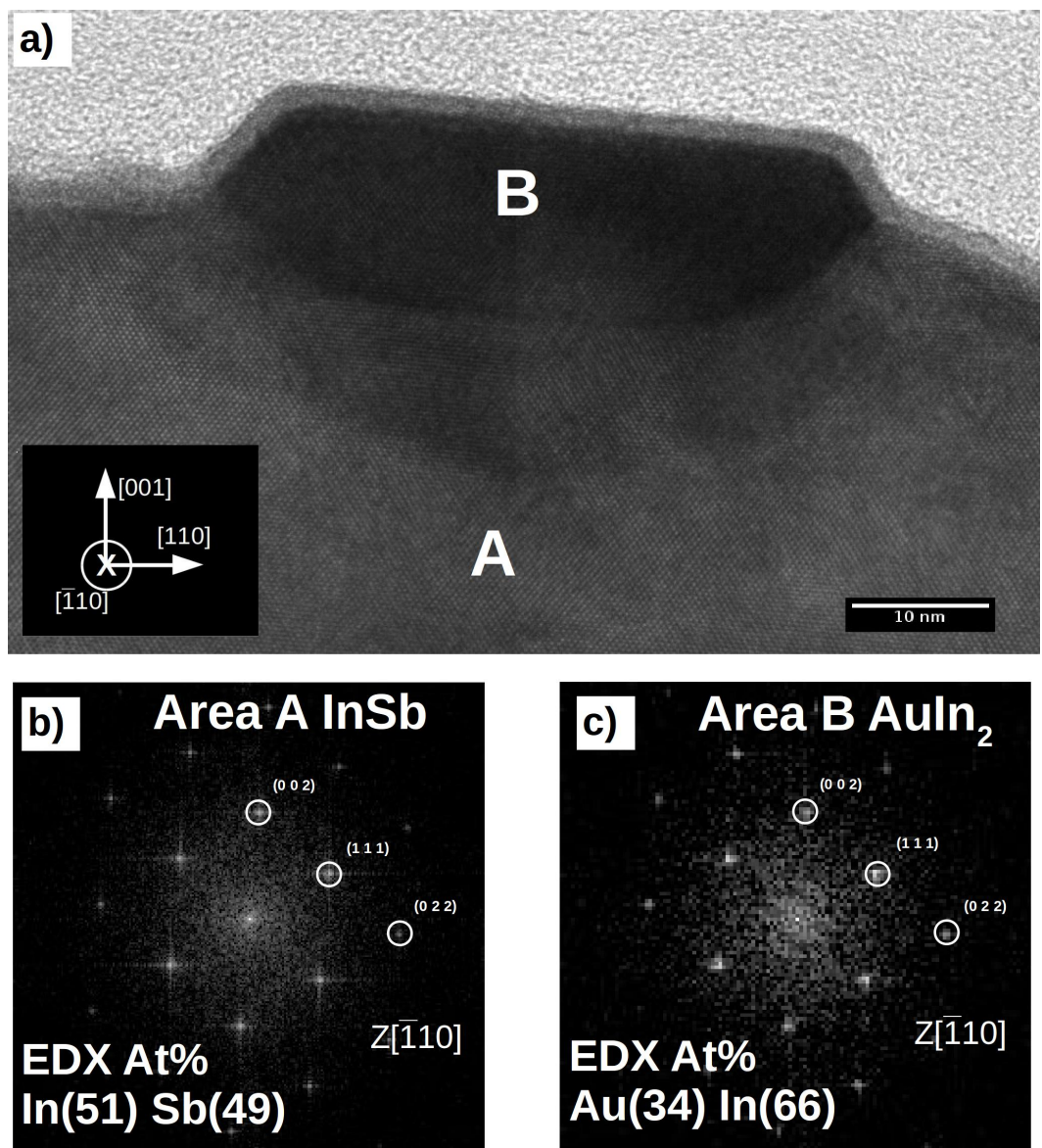

Figure S4: a) HRTEM of AuIn2 nanowire cross-section grown on InSb(001) surface. Indexed FFT spots and results of EDX analysis of InSb area A b) and AuIn2 nanowire area B c). Crystallographic direction of InSb indicated.

# Monte-Carlo simulation of electrons interaction by CASINO for Au nanostructures on Germanium

## Pure Germanium

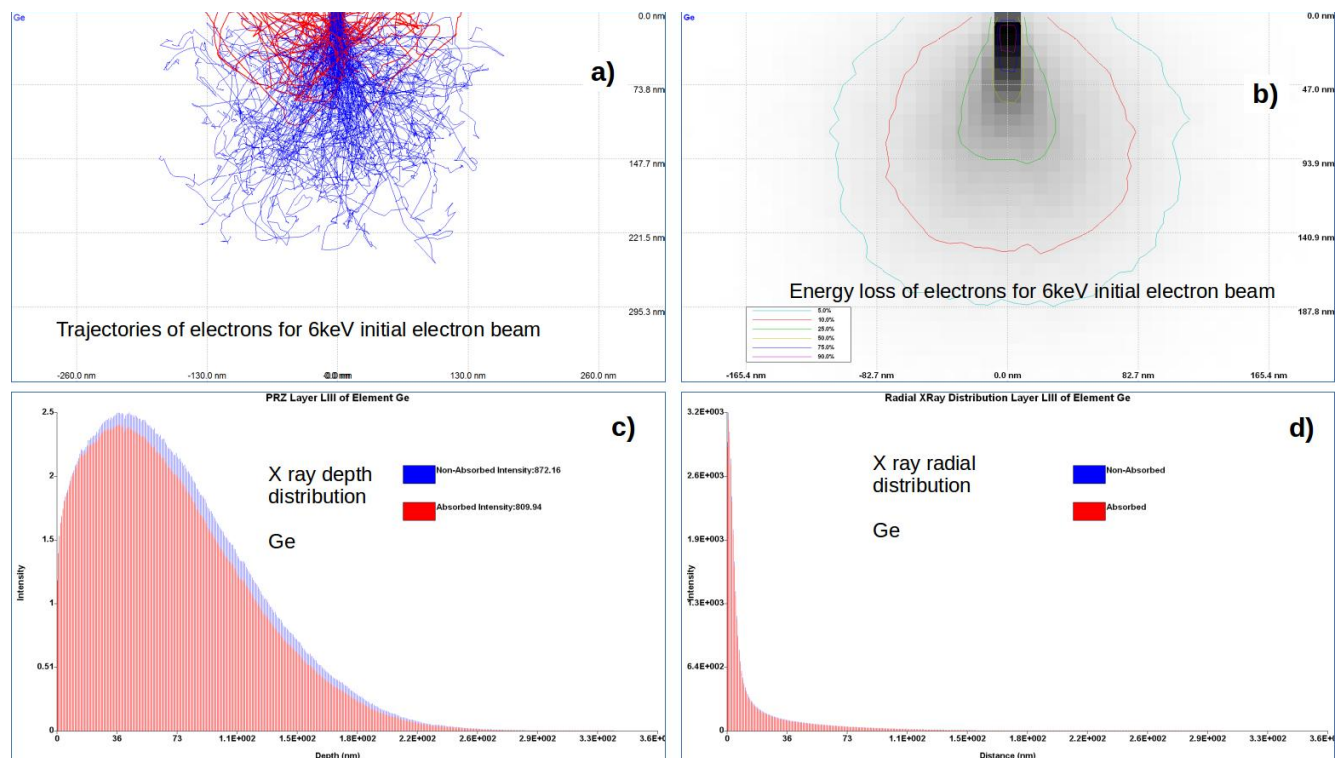

*Figure S5: Monte-Carlo simulation of 6keV electrons interaction with Germanium by CASINO. a) Trajectories of generated electrons, b) energy loss of electrons. c) generated X rays depth distribution. d) generated X rays radial distribution.*

# Au nanostructures on Germanium

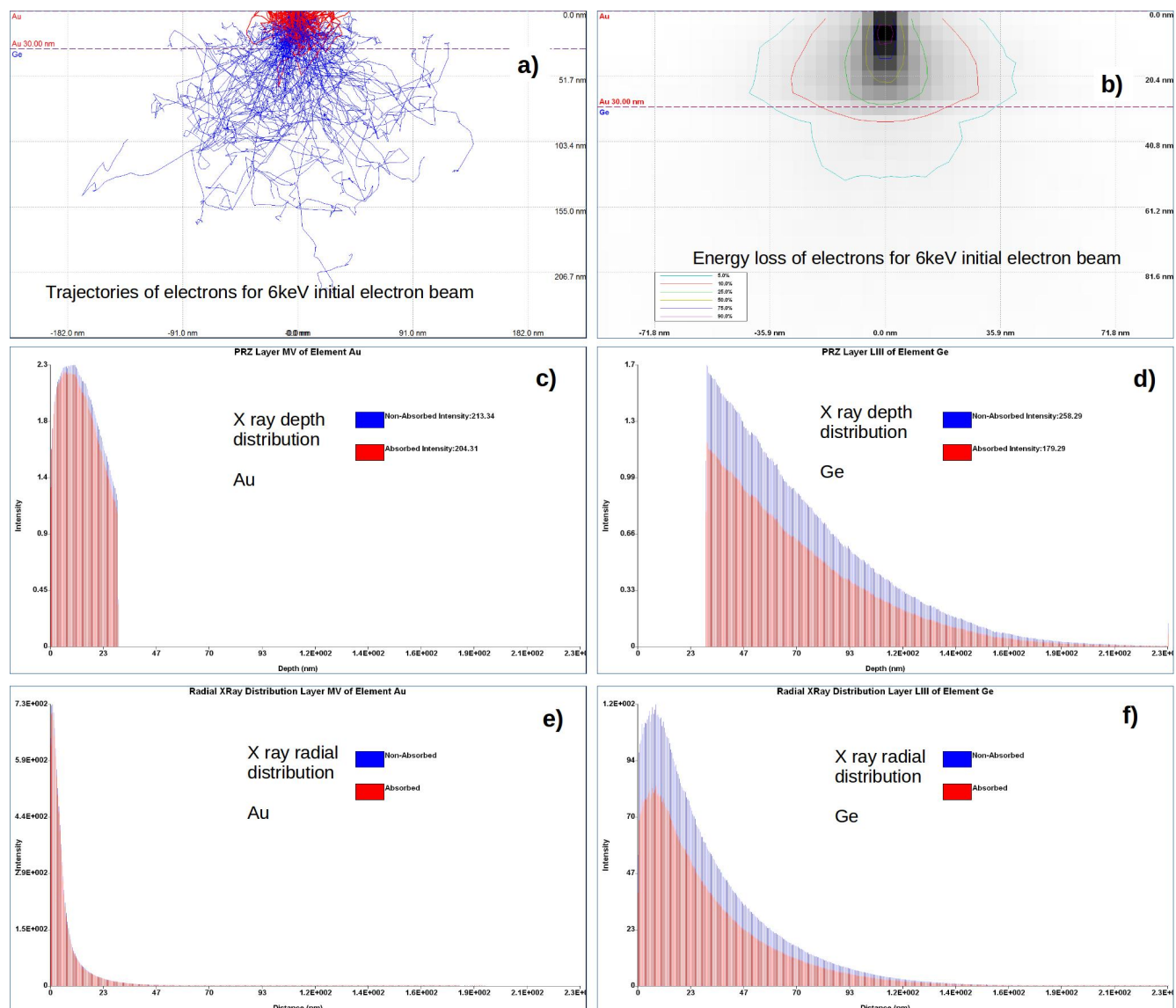

*Figure S6: Monte-Carlo simulation of 6keV electrons interaction with Au nanostructure(30nm) on Germanium by CASINO. a) Trajectories of generated electrons, b) energy loss of electrons. Generated X rays depth distribution for gold c) and germanium d). Generated X rays radial distribution for gold e) and germanium f).*